\documentclass[preprintnumbers,amsmath,11pt,amssymb,floatfix,superscriptaddress,nofootinbib]{revtex4}
\usepackage{graphicx}
\usepackage{epsfig}
\usepackage{bm}
\usepackage{amsfonts}

\begin{document}

\title{\Large Can Holographic dark energy increase the mass of the wormhole?}

\author{Surajit Chattopadhyay}
\affiliation{ Department of Computer Application, Pailan College
of Management and Technology, Bengal Pailan Park, Kolkata-700 104,
India.} \footnote{$^{,}$$^{*}$ Corresponding author, email:
surajitchatto@outlook.com, surajcha@iucaa.ernet.in}

\author{Davood Momeni}
\email{d.momeni@yahoo.com}
\affiliation{Eurasian International Center for Theoretical Physics and Department of General
Theoretical Physics, Eurasian National University, Astana 010008, Kazakhstan.}

\author{Aziza Altaibayeva}
\email{aziza.bibol@mail.ru}
\affiliation{Eurasian International Center for Theoretical Physics and Department of General
Theoretical Physics, Eurasian National University, Astana 010008, Kazakhstan.}

\author{Ratbay Myrzakulov}
\email{rmyrzakulov@gmail.com}
\affiliation{Eurasian International Center for Theoretical Physics and Department of General
Theoretical Physics, Eurasian National University, Astana 010008, Kazakhstan.}

\begin{abstract}
\textbf{Abstract:} In this work, we have studied accretion of dark energy (DE) onto Morris- Thorne wormhole with three different forms, namely, holographic dark energy, holographic Ricci dark energy and modified holographic Ricci dark energy . Considering the scale factor in power-law form we have observed that as the holographic dark energy accretes onto wormhole, the mass of the wormhole is decreasing. In the next phase we considered three parameterization schemes that are able to get hold of quintessence as well as phantom phases. Without any choice of scale factor we reconstructed Hubble parameter from conservation equation and dark energy densities and subsequently got the mass of the wormhole separately for accretion of the three dark energy candidates. It was observed that if these dark energies accrete onto the wormhole, then for quintessence stage, wormhole mass decreases up to a certain finite value and then again increases to aggressively during phantom phase of the universe.
\end{abstract}


\maketitle

\section{Introduction}
Accelerated expansion of our universe (see \cite{obs1,obs2,obs3}), as evidenced by Supernovae Ia (SNeIa),
Cosmic Microwave Background (CMB) radiation anisotropies, Large Scale Structure (LSS) and X-ray experiments, has
stimulated interest in models containing a component with an arbitrary equation of state $w=p/\rho\leq-1$  where $\rho$ is the dark energy density and $p$ is the pressure. Such a component is dubbed as ``quintessence" \cite{quint1}. DE is proposed as a fluid with a negative efective pressure $p$ and energy density $\rho$ satisfies $p<-\rho$. It violates  the weak energy condition \cite{farao}. So, it plays the role of an exotic fluid ( see for recent reviews   \cite{rev1,rev2,rev3,rev4,rev5} ).\par
Acceration of the DE on a black hole is a phenomena which it changes (decreases or increases?)  the mass of the black hole \cite{Babichev}. Not only black holes \cite{ujjal2,jam,jam1} but other black objects has the same physical scenarion like 
the acceration of DE on to  wiggly cosmic strings \cite{Gonz´alez} and specially the geometry of a wormholes \cite{wh1,wh2,ujjal1,pedro}. \par
Wornholes has been proposed to be a geometry of a bridge which is able to connect two Lorentzian metrics of the spacetime in a causual way.
Wormholes are expected usually due to quantum effects \cite{gr-qc/9904035,hep-th/9812164}. They inspired from the quantum effects \cite{hep-th/0502082,Jamil:2011iu}.
 Different aspects of wormholes are widely studied in literature from general relativity to modified gravities  \cite{ujjal1}-\cite{Jamil:2012ti}. Wormholes can be immersed in 
in a Friedmann–Lemaitre–Robertson–
Walker (FLRW) universe  \cite{wh1,P.F.,Nuovo Cimento B}. In such cases it has been showed that  the accretion of a certain kinds of DE, phantom energy  can make the wormhple massive  . So, it works as an stabilizer of the wormhole. Such acceration phenomena has been investigated in literature  \cite{jamil2,jamil3}. Different scenarios proposed for phantom DE like 
 generalised
Chaplygin gas model (PGCG)  \cite{podaro}.

In the present study, we have considered accretion of holographic dark energy (HDE) with density $\rho_{HDE}=3c^2M_p^2H^2$ \cite{HDE1,HDE2,HDE3,HDE4}; holographic Ricci dark energy with density $\rho_{HRDE}=3c^2(\dot{H}+2H^2)$ \cite{RDE1,RDE2,RDE3}, and modified holographic Ricci dark energy (MHRDE) with density  $\rho_{MHRDE}=\frac{2}{\alpha-\beta}\left(\dot{H}+\frac{3\alpha}{2}H^2\right)$ ($\alpha$ and $\beta$ are free constants) \cite{MHRDE1,MHRDE2,MHRDE3}
  on Morris-Thorne Wormhole \cite{morris}(for a more motivated and physically important model see \cite{GHDE}). This study is primarily motivated by considering two types of Chaplygin gas model accreting onto Morris-Thorne Wormhole and  different parameterization schemes and increasing of the  wormhole mass  during the phantom phase of the universe and decreasing during quintessence.  It was understood that wormholes
would shrink if Hubble parameter $H$ decreases, we decided to select power-law form of the scale factor leading to decreasing $H$ and then investigated accretions of three dark energy models, whose densities are functions of $H$\cite{podaro}.

Plan of the paper is as follows. In section II we shall give a brief overview of Morris-Thorne Wormhole, in section III we shall discuss accretion of the dark energy candidates on it and in section IV, we shall discuss various parameterization schemes and finally we shall conclude in section IV.

\section{A brief overview of Morris-Thorne Wormhole}
Wormholes may be classified into two categories \cite{ujjal1}: Euclidean wormholes and Lorentzian wormholes. The Euclidean wormholes arise in Euclidean quantum gravity and the Lorentzian wormholes, which are static spherically symmetric solutions of Einstein’s general relativistic field equations. The Lorentzian wormhole consists of two asymptotically flat spacetimes and the bridge connecting two space-times. In general, the flatness of two spacetimes is not necessarily required for constructing wormholes.  Morris-Thorne solution  is the most important model of simplicity \cite{worm1}however is not the unique static solution \cite{worm2,worm3,worm4,worm5,worm6,worm7}.

Non-static spherically symmetric Morris-Thorne wormhole metric is given by \cite{morris}
\begin{equation}\label{metric}
ds^2=-e^{\Phi(r)}dt^2+\frac{dr^2}{1-\frac{K(r)}{r}}+r^2(d\theta^2+\sin^2\theta d\phi^2)
\end{equation}
Here  $K(r)$ is shape function and $\Phi(r)$ is redshift function. Also we denote by $r$ the radial coordinate of a co-moving observer . If $K(r_0)=r_0$, the radius $r_0$ is called the wormhole throat radius. As we are considering the phenomenon of dark energy accretion onto wormhole, we shall consider $r_0< \infty$.
In the present accretion problem we shall follow the procedure of \cite{Babichev} . So, we consider the tensor momentum-energy for a relativistic perfect fluid
\begin{equation}\label{tensor}
T_{\mu\nu}=(p+\rho)u_{\mu}u_{\nu}+pg_{\mu\nu}
\end{equation}
in which $p$ is pressure, $\rho$ is energy density and $u^{\mu}=dx^{\mu}/ds$ is the four-velocity satisfying $u^{\mu}u_{\mu}=-1$. We ignore of backreactions on the static wormhole background  and it is assumed $\Phi(r)=0$. The relativistic Bernoulli’s equation after the time component of the energy-momentum conservation law $u_{\mu}T_{;\nu}^{\mu\nu}=0$ provide the first integral of motion for static, spherically symmetric accretion onto wormhole and it yields ($u=\frac{dr}{ds}>0$ )\cite{pedro}
\begin{equation}\label{int1}
M^{-2}r^2u(p+\rho)\left(1-\frac{K(r)}{r}\right)^{-1}\left(u^2+\frac{K(r)}{r}-1\right)^{\frac{1}{2}}=C_1
\end{equation}
where $M$ is the exotic mass of the wormholeand we introduced  the integration constant $C_1$ .  An auxiliarly equation is obtained by the projection of the energy-momentum tensor on the four-velocity $u_{\mu}T^{\mu\nu}_{;\nu}=0$.  In the case of the relativistic  perfect fluid and in the background of a static wormhole, one finally obtains
\begin{equation}
M^{-2}r^2u\left(1-\frac{K(r)}{r}\right)^{-1}\exp\left[\int_{\rho_{\infty}}^{\rho}\frac{d\rho}{\rho+p(\rho)}\right]=C_2
\end{equation}
where, where$C_2>0$ is dimensionless integration constant. We assume that wormhole is asymptotically flat with a zero angular momentum (ZAMO) observer at infinity. So, the magnitude of the energy density of DE at infinity is denoted by  $\rho_{\infty}$ .

The rate of change of the exotic wormhole mass $\dot{M}$ is computed as the follows \cite{pedro} :
\begin{equation}
\dot{M}=\int dS T_0^{r}
\end{equation}
Using the above results we can obtain the rate of change of the gravitational mass as
\begin{equation}
\dot{M}=-4\pi M^2 Q\sqrt{1-\frac{K(r)}{r}}(p+\rho)
\end{equation}
with the constant $Q>0$. For the asymptotic regime $r\rightarrow\infty$, the rate $\dot{M}$ reduces to
\begin{equation}\label{Mdot}
\dot{M}=-4\pi M^2Q(p+\rho)
\end{equation}
Clearly, if $\dot{H}<0$, then $\dot{M}<0$. To investigate increasing or decreasing in mass function we should solve (\ref{Mdot}) to find $M(t)$. It depends on how $\{p,\rho\}$ change as functions of time.

\section{Accretion of holographic dark energy on Morris-Thorne Wormhole}
Power-law scale factor is examined by observational data as a good and physically reasonable assumption\cite{Rani:2014sia}. One reason is here we have only two fitting parameters , the Hubble parameter is evaluated at the present time $H_0$ and the value of decelation parameter $q=-1-\frac{\dot{H}}{H^2}$ computed at the present time $q_0$. 
We assume the scale factor in power-law form as
\begin{equation}\label{a}
a(t)=a_0 t^m,~~m>0
\end{equation}
It is well understood that the model described by the above equation for $m>1$ is able to address the horizon,flatness and age problem of the early universe \cite{early1,early2,early3}. By a very carefully fitting method, we observe that the power law model modelis preferred over other models of $a(t)$  if we use a  ﬁtting  procedure with only two parameters which it is here $\{H_0,q_0\}$. Power-law scale factors have been used in different modified gravities models \cite{Bamba:2014mya,Bamba:2013fha}
\par
In our study on acceration of holographic dark energy on wormholes,we use this power-law solution. For the power law, the Hubble's parameter reads:
\begin{equation}\label{H}
H=\frac{m}{t}
\end{equation}
Since we are considering 
three dark energy candidates ,we list them here for the power-law \\
Holographic dark energy \cite{HDE1}
\begin{equation}\label{hde}
\rho_{HDE}=\frac{3 c^2 M_p^2 m^2}{t^2}
\end{equation}
Holographic Ricci dark energy \cite{RDE1}
\begin{equation}\label{hrde}
\rho_{HRDE}=\frac{3 c^2 (-m+2m^2)}{t^2}
\end{equation}
Modified holographic Ricci dark energy \cite{MHRDE1}
\begin{equation}\label{mhrde}
\rho_{MHRDE}=\frac{2 \left(-m+\frac{3 m^2 \alpha }{2 }\right)}{(\alpha -\beta)t^2 }
\end{equation}
To proceed the acceration program we need to specify $\{\rho(t),p(t)\}$. Using the first and second FRW equations :
\begin{eqnarray}
H^2=\frac{1}{3}\rho\\
\dot{H}=-\frac{1}{2}(\rho+p)
\end{eqnarray}
and by remembering the energy conservation equation:
\begin{eqnarray}
\dot{\rho}+3H(\rho+p)=0
\end{eqnarray}
 we can get the pressure for the above dark energies
\begin{equation}\label{phde}
p_{HDE}=-\frac{c^2 (2+3 m) M_p^2 t^2}{m^3}
\end{equation}
\begin{equation}\label{phrde}
p_{HRDE}=\frac{c^2 \left(-2+7 m-6 m^2\right)}{t^2}
\end{equation}
\begin{equation}\label{pmhrde}
p_{MHRDE}=-\frac{(-2+3 m) (-2+3 m \alpha )}{3 t^2 (\alpha -\beta )}
\end{equation}

Using Eqs. (\ref{hde}) and (\ref{phde}) in (\ref{Mdot}) (i.e. for HDE)
\begin{equation}\label{Mhdedot}
\dot{M}=-\frac{8 M^2 \pi  Q c M_p m}{t^2}
\end{equation}
Using Eqs. (\ref{hrde}) and (\ref{phrde}) in (\ref{Mdot}) (i.e. for HRDE)
\begin{equation}\label{Mhrdedot}
\dot{M}=-\frac{8 M^2 \pi  Q c\sqrt{ m (-1+2 m)}}{t^2}
\end{equation}
Using Eqs. (\ref{mhrde}) and (\ref{pmhrde}) in (\ref{Mdot}) (i.e. for MHRDE)
\begin{equation}\label{Mmhrdedot}
\dot{M}=-\frac{8 M^2 \pi  Q \sqrt{\frac{m (-2+3 m \alpha )}{ (\alpha -\beta )}}}{\sqrt{3} t^2}
\end{equation}
Solving Eqs. (\ref{Mhdedot}), (\ref{Mhrdedot}) and (\ref{Mmhrdedot}) we get respectively
\begin{equation}\label{Mhde}
M_1=\left(\frac{1}{M_0}+8c m M_p \pi Q\left(H_0-a_0(1+z)^m\right)\right)^{-1}~~~\textrm{(for HDE)}
\end{equation}
\begin{equation}\label{Mhrde}
M_2=\frac{\left(\frac{1}{a_0 (1+z)}\right)^{\frac{1}{m}}}{-8 c \sqrt{m} \sqrt{-1+2 m} \pi  Q-\frac{\left(-8 c \sqrt{m} \sqrt{-1+2 m}
M_0 \pi  Q-t_0\right) \left(\frac{1}{a_0 (1+z)}\right)^{\frac{1}{m}}}{M_0 t_0}}~~~\textrm{(for HRDE)}
\end{equation}
\begin{equation}\label{Mmhrde}
M_3=\frac{3 \left(\frac{1}{a_0 (1+z)}\right)^{\frac{1}{m}} \sqrt{\alpha -\beta }}{-8 \sqrt{3} \sqrt{m} \pi  Q \sqrt{-2+3 m \alpha }-\frac{\left(\frac{1}{a_0
(1+z)}\right)^{\frac{1}{m}} \left(-3 t_0-\frac{8 \sqrt{3} \sqrt{m} M_0 \pi  Q \sqrt{-2+3 m \alpha }}{\sqrt{\alpha -\beta }}\right) \sqrt{\alpha
-\beta }}{M_0 t_0}}~~~\textrm{(for MHRDE)}
\end{equation}

In the above equations $M_0$ and $H_0$ are mass of wormhole and value of the Hubble parameter respectively at $t=t_0$ and we have used $t=\left[a_0 (1+z)\right]^{-1/m}$.

It is adequate to use a Taylor series expansion for redshift $z\in [-2,3]$. For this reason we remembering to mind that:
\begin{eqnarray}
\frac{1}{(1-x)^{\alpha}}=\Sigma_{n=0}^{\infty}\frac{(\alpha+n-1)!}{n!(\alpha-1)!}x^n,\ \ |x|<1
\end{eqnarray}
Using this series for $\alpha=\frac{1}{m},x=-z$ we obtain:
\begin{eqnarray}\label{taylor}
(1+z)^{-1/m}=(-1)^{-1/m}+\frac{(-1)^{-1/m} (z+2)}{m}+\frac{(-1)^{-1/m} (1+m) (z+2)^2}{2 m^2}+\nonumber\\
~~~~~~~~~~~~~~~~~~~~~~~~~~~~~~~~~~~~~~~~~~~~~~~\frac{(-1)^{-1/m} (1+m) (1+2 m) (z+2)^3}{6 m^3}+O[z+2]^4
\end{eqnarray}
Using the above expansion (\ref{taylor}) we have at very late stage i.e. $z\rightarrow~-1$ the mass of wormhole as HDE, HRDE and MHRDE accretion occurs as follows:
\begin{eqnarray}
M_1=\left[\frac{1}{M_0}+8 c m \left(H_0-6^{-1/m} \left(\frac{(-1)^{\frac{1}{m}} m^3}{a_0 \left(1+6 m+11 m^2+6 m^3\right)}\right)^{-1/m}\right)
M_p \pi  Q\right]^{-1}\label{M1}
\end{eqnarray}
\begin{equation}
\begin{array}{c}\label{M2}
M_2=\left(9 (-a_0)^{2/m} m^{11/2} M_0\right)\times\left[2 c H_0 \sqrt{-1+2 m} M_0 \pi  Q+m \left(9 (-a_0)^{2/m} m^{9/2}+\right.\right.\\
\left.\left.24
c H_0 \sqrt{-1+2 m} M_0 \pi  Q+116 c H_0 m \sqrt{-1+2 m} M_0 \pi  Q-12 c \left((-a_0)^{\frac{1}{m}}-24 H_0\right)\right.\right.\\
\left.\left.m^2 \sqrt{-1+2 m} M_0 \pi  Q-2 c \left(36 (-a_0)^{\frac{1}{m}}-193 H_0\right) m^3 \sqrt{-1+2 m} M_0 \pi  Q-132 c \left((-a_0)^{\frac{1}{m}}-\right.\right.\right.\\
\left.\left.\left.2 H_0\right) m^4 \sqrt{-1+2 m} M_0 \pi  Q-72 c \left((-a_0)^{\frac{1}{m}}-H_0\right) m^5 \sqrt{-1+2 m} M_0 \pi  Q\right)\right]^{-1}
\end{array}
\end{equation}
\begin{equation}
\begin{array}{c}\label{M3}
M_3=\left(\frac{1}{M_0}-\frac{8 \sqrt{m} \pi  Q \left(-1+m \left(-6+m \left(-11+6 m \left(-1+(-a_0)^{\frac{1}{m}} t_0\right)\right)\right)\right)
\sqrt{-2+3 m \alpha }}{\sqrt{3} (1+m) (1+2 m) (1+3 m) t_0 \sqrt{\alpha -\beta }}\right)^{-1}
\end{array}
\end{equation}

\begin{figure}[ht] \begin{minipage}[b]{0.45\linewidth} \centering\includegraphics[width=\textwidth]{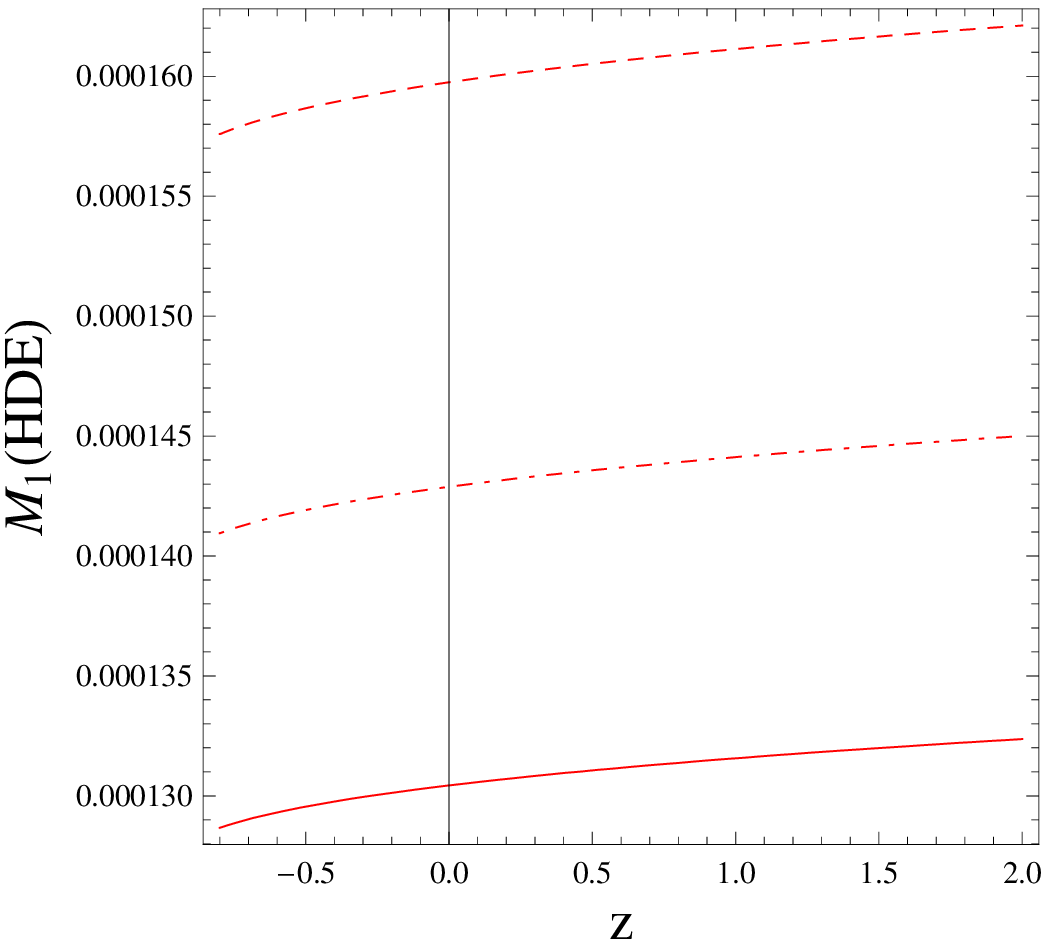} \caption{Variations of wormhole mass $M_1$ (Eq. (\ref{M1})) against redshift $z=a^{-1}-1$ for HDE.} \label{M1plot} \end{minipage} \hspace{0.5cm} \begin{minipage}[b]{0.45\linewidth}
\centering\includegraphics[width=\textwidth]{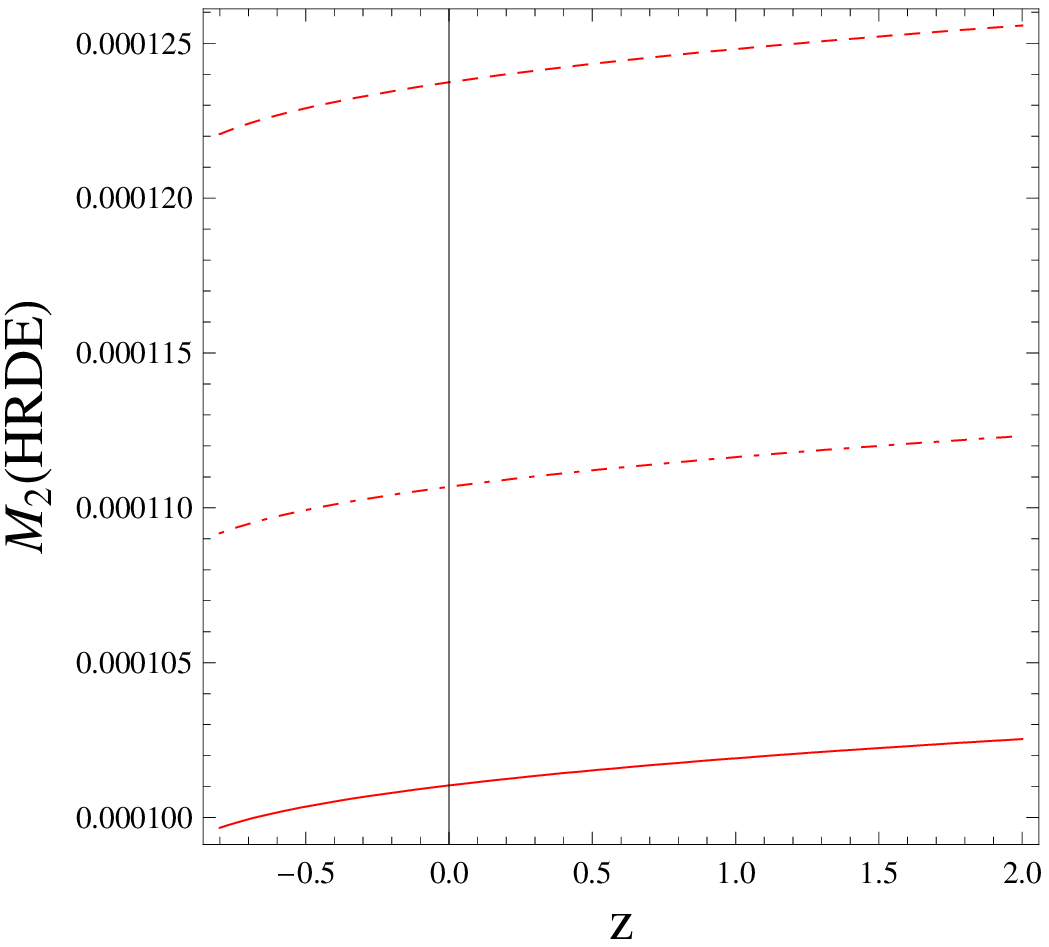} \caption{Variations of wormhole mass $M_2$ (Eq. (\ref{M2})) against redshift $z=a^{-1}-1$ for HRDE.}
\label{M2plot} \end{minipage} \hspace{0.5cm} \begin{minipage}[b]{0.45\linewidth}
\centering\includegraphics[width=\textwidth]{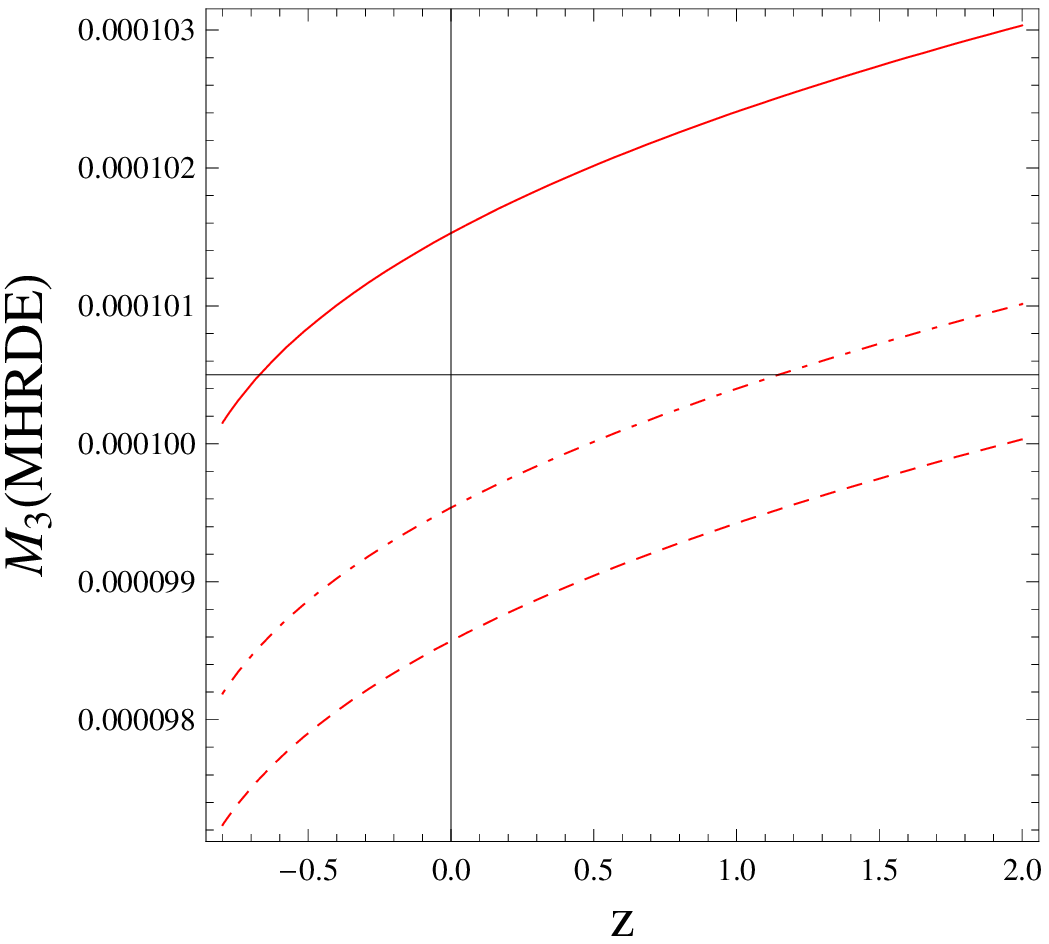} \caption{Variations of wormhole mass $M_3$ (Eq. (\ref{M3})) against redshift $z=a^{-1}-1$ for MHRDE.}
\label{M3plot} \end{minipage}  \end{figure}
In Figures (\ref{M1plot},\ref{M2plot},\ref{M3plot}) we plot the mass functions versus red shift $z$. 
 As the accretion of HDE occurs on the wormhole (Fig. \ref{M1plot}), its mass is increasing. 
Fig. \ref{M2plot} shows that for accretion of HRDE the mass of wormhole shows increasingpattern. Fig. \ref{M3plot} shows similar pattern to that of HRDE and here also the accretion of MHRDE onto wormhole leads to  the mass increases indefinitely.

\section{Parameterizations of dark energy Models}
The main important parameter in DE scenario is the equation of state (EoS) parameter $w$. It is defined as the ratio between the effective pressure of DE to the energy density. In general,EoS is function of cosmological time $w(t)$. But a better parametrization is obtained when we write it as function of redshift $w(z)$. So, we have
 $w(z)=\frac{p(z)}{\rho(z)}$. There is no direct way to analyze and find this function in a fully theoretical based formalism. For low redshifts,near the now era we have two known families of models described by "Families I,II". In the common forms we have:
\begin{enumerate}
  \item Family I: $w(z)=w_0+w_1\left(\frac{z}{1+z}\right)^m$
  \item Family II:$w(z)=w_0+w_1\frac{z}{(1+z)^m}$
\end{enumerate}
The strategy is to insert this $w(z)$ in the FRW and continuty equations to find $\{\rho(z),p(z)\}$ . The benefit is in this reconstruction scheme,there is no need to assume any form of scale factor. \par
For family I, we get from the conservation equation that
\begin{equation}\label{rhopara1}
\rho(z)=\rho_0 e^{(-1)^{-m} 3 w_1  \text{Beta}[1+z,-m,1+m]} (1+z)^{3(1+ w_0)}
\end{equation}
For family II, we get from the conservation equation that
\begin{equation}\label{rhopara2}
\rho(z)=\rho_0 e^{-\frac{3 w_1 \left(-1+\left(\frac{1}{1+z}\right)^m (1+m z)\right)}{(-1+m) m}}  (1+z)^{3(1+3 w_0)}
\end{equation}
where, $w_0$ and $w_1$ are two unknown parameters, which can be constrained by the recent observations and $m$ is
a natural number.

In this section we shall not assume any form of scale factor. Wee shall consider the mass function in terms of the Hubble parameter $H$ as in the form \cite{pedro,ujjal1}
\begin{equation}\label{mass}
M=\frac{M_0}{1-8\pi Q M_0(H-H_0)}
\end{equation}
By considering the form of $H(z)$ we can check whether $M(z)$ is an increasing or decreasing function.

\subsection{CPL parameterization}
\begin{figure}[h]
\begin{minipage}{16pc}
\includegraphics[width=16pc]{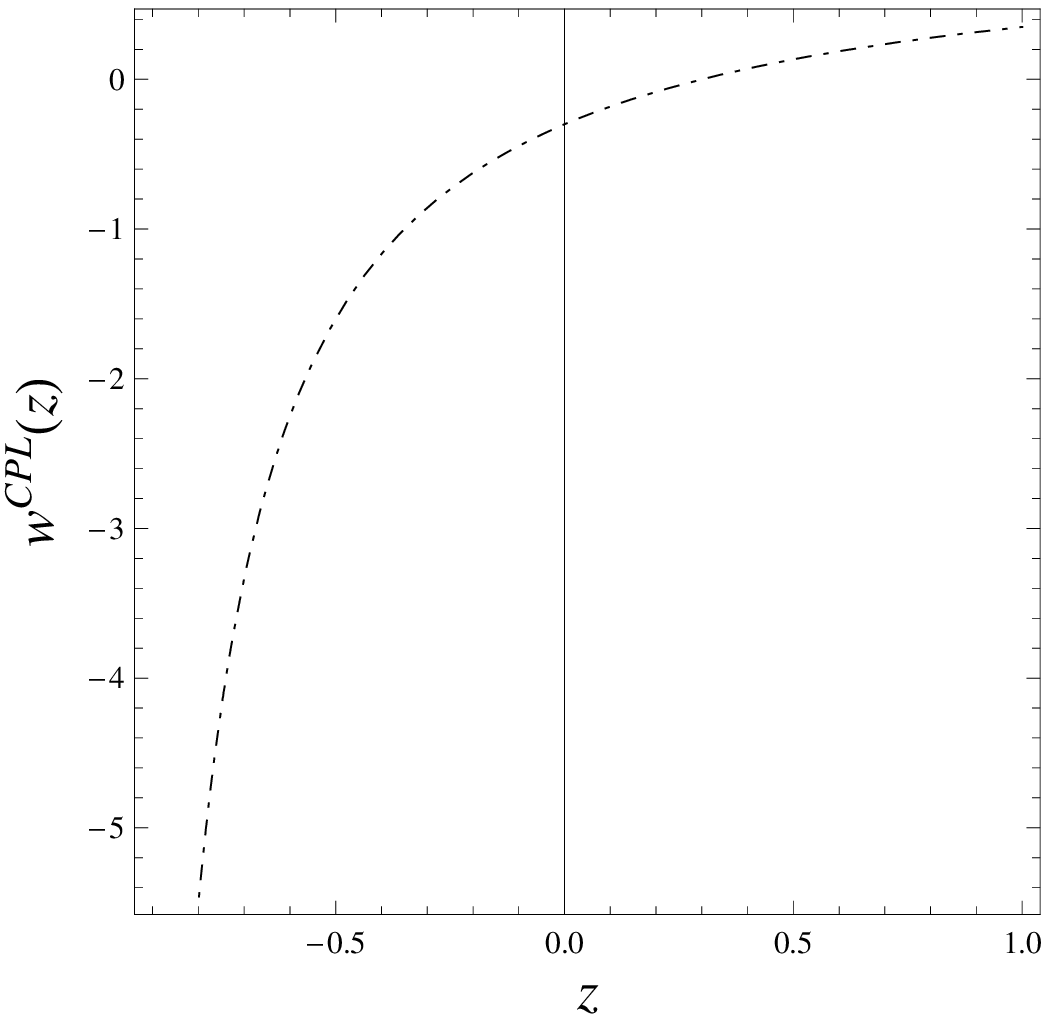}
\caption{\label{fig:eoscpl} EoS for JBP parameterization with $w_0=-0.3,~w_1=1.3$.}
\end{minipage}\hspace{3pc}%
\begin{minipage}{16pc}
\includegraphics[width=16pc]{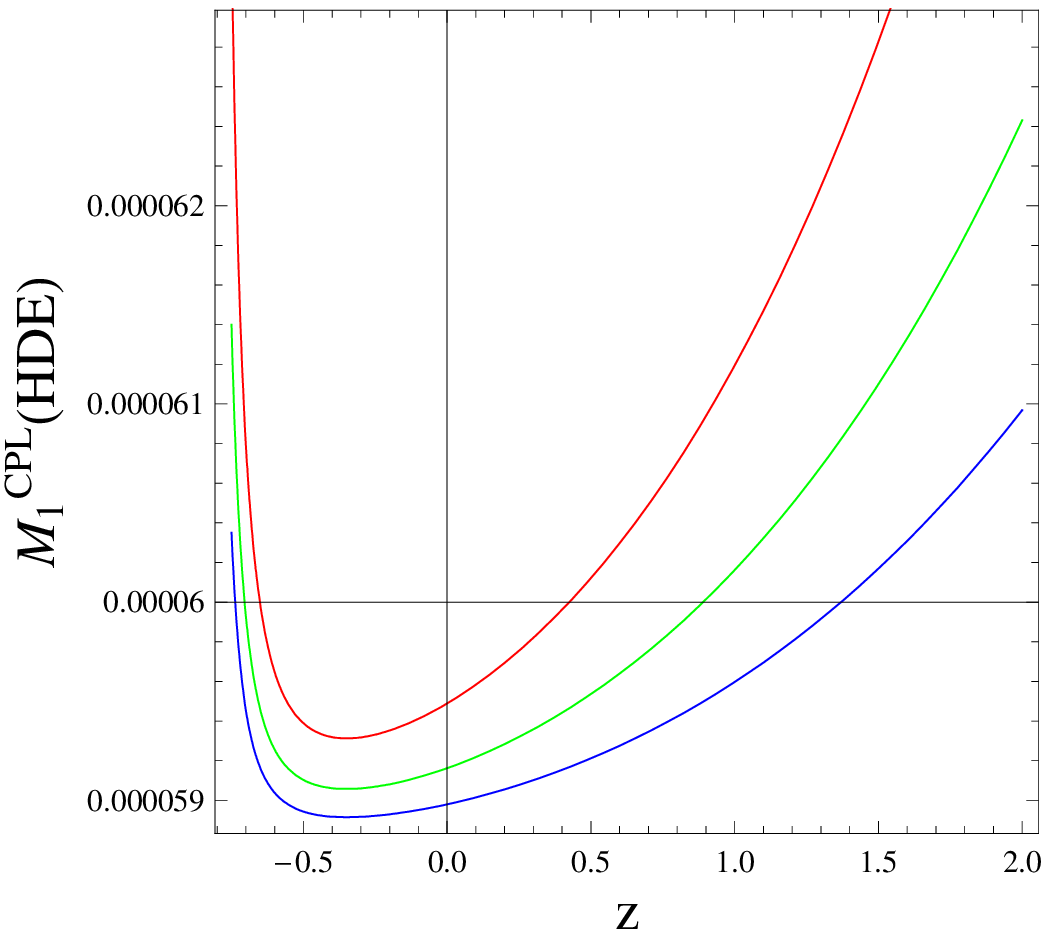}
\caption{\label{fig:M1cpl} Variations of wormhole mass $M_1$ (Eq. (\ref{M1})) against redshift $z=a^{-1}-1$ for HDE using CPL parameterization.}
\end{minipage}\hspace{3pc}%
\begin{minipage}{16pc}
\includegraphics[width=16pc]{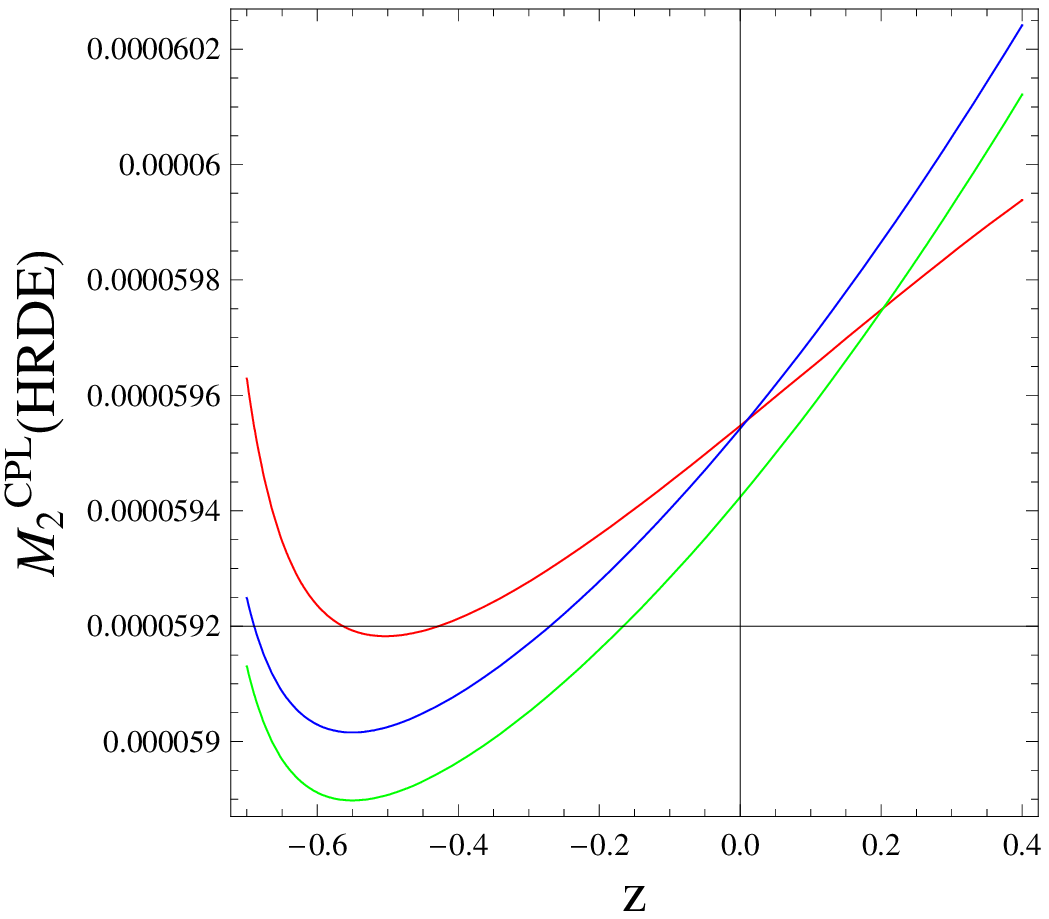}
\caption{\label{fig:M2cpl}Variations of wormhole mass $M_2$ (Eq. (\ref{M2})) against redshift $z=a^{-1}-1$ for HRDE using CPL parameterization.}
\end{minipage}\hspace{3pc}%
\begin{minipage}{16pc}
\includegraphics[width=16pc]{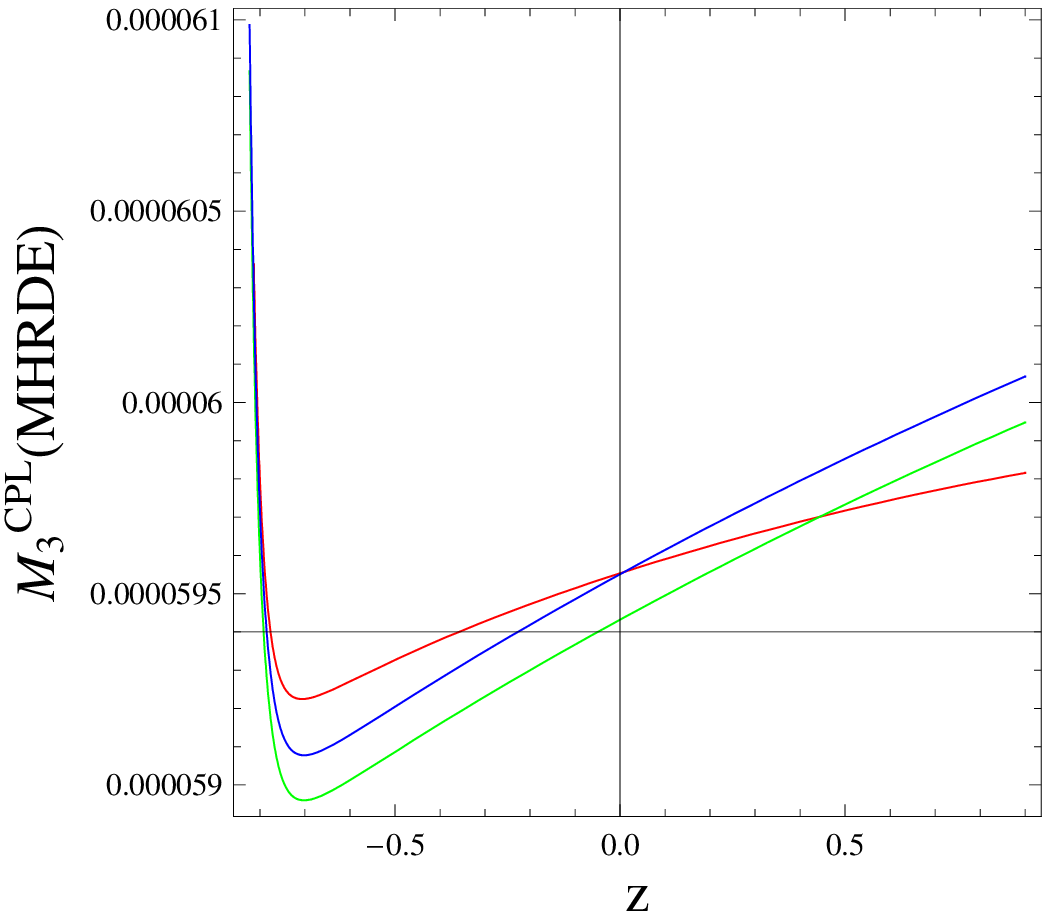}
\caption{\label{fig:M3cpl}Variations of wormhole mass $M_3$ (Eq. (\ref{M3})) against redshift $z=a^{-1}-1$ for MHRDE using CPL parameterization.}
\end{minipage}\hspace{3pc}%
\end{figure}

A popular parametrization, proposed by Chevallier and Polarski \cite{CPL1} and Linder (CPL) \cite{CPL2}, explains evolution of dark energy in a flat FRW Universe. It is termed as the CPL model and it proposes the following parameterization
of the EoS:
\begin{equation}\label{CPL}
w^{CPL}(z)=w_0+w_1 \frac{z}{1+z}
\end{equation}
We get $w^{CPL}$ if we put $m=1$ in Family I as well as Family II. The conservation equation takes the form
\begin{equation}
(1+z) \frac{d\rho}{dz}=3 \rho \left(1+w_0+\frac{w_1 z}{1+z}\right)
\end{equation}
and solving the above we get
\begin{equation}\label{rhocpl}
\rho=\rho_0 e^{-\frac{3 w_1z}{1+z}} (1+z)^{3 (1+w_0+w_1)}
\end{equation}
Considering $\rho=\rho_{HDE}=3c^2M_p^2H^2$ we get
\begin{equation}\label{hsqrHDE}
H^2=\frac{1}{3 c^2 M_p^2}e^{-\frac{3 e_1 z}{1+z}} \rho_0 (1+z)^{3 (1+w_0+w_1)}
\end{equation}
Using Eq. (\ref{hsqrHDE}) in (\ref{mass}) we get the mass $M$ of wormhole as a function of $z$ and plot it in Fig. \ref{fig:M1cpl}.  Solving the differential equation coming out of the consideration $\rho=\rho_{HRDE}=3c^2(\dot{H}+2H^2)$ we get
\begin{eqnarray}\label{hsqrHRDE}
H^2=\left(e^{-3 w_1} (1+z)^4 \left(9 c^2 e^{3 w_1} H_0^2 w_1-2\times3^{3 (w_0+w_1)} \rho_0 (-w_1)^{3
(w_0+w_1)} \Gamma[1-3 w_0-3 w_1,-3 w_1]+\right.\right.\nonumber\\
\left.\left.2\times3^{3 (w_0+w_1)} \rho_0 \left(-\frac{w_1}{1+z}\right)^{3
(w_0+w_1)} (1+z)^{3 (w_0+w_1)} \Gamma\left[1-3 w_0-3 w_1,-\frac{3 w_1}{1+z}\right]\right)\right)\times
(9 c^2
w_1)^{-1}\nonumber\\
\end{eqnarray}
Using Eq. (\ref{hsqrHRDE}) in (\ref{mass}) we get the mass $M$ of wormhole as a function of $z$ and plot it in Fig. \ref{fig:M2cpl}. Solving the differential equation coming out of the consideration $\rho=\rho_{MHRDE}=\frac{2}{\alpha-\beta}\left(\dot{H}+\frac{3\alpha}{2}H^2\right)$ we get
\begin{eqnarray}\label{hsqrMHRDE}
&&H^2=H_0^2+3^{3 (1+w_0+w_1-\alpha )} e^{-3 w_1}\rho_0 (-w_1)^{3 (1+w_0+w_1-\alpha )} (\alpha
-\beta ) \Gamma[-3 (1+w_0+w_1-\alpha ),-3 w_1]+\nonumber\\&&
3^{3 (1+w_0+w_1-\alpha )} e^{-3 w_1}\rho_0 w_1^3
\left(-\frac{w_1}{1+z}\right)^{3 (w_0+w_1-\alpha )} (1+z)^{3 (w_0+w_1)} (\alpha -\beta )\times\nonumber\\&&
 \Gamma\left[-3 (1+w_0+w_1-\alpha
),-\frac{3 w_1}{1+z}\right]\nonumber\\&&
\end{eqnarray}
Using Eq. (\ref{hsqrMHRDE}) in (\ref{mass}) we get the mass $M$ of wormhole as a function of $z$ and plot it in Fig. \ref{fig:M3cpl}.

In Fig. \ref{fig:eoscpl} we have plotted the EoS parameter for $w_0=-0.3,~w_1=1.3$ and observed a transition from quintessence to phantom phase at $z\approx-0.4$. As the accretion of HDE occurs on the wormhole (Fig. \ref{fig:M1cpl}), its mass is decreasing as long as the universe is in quintessence phase and as soon as it enters the phantom phase the mass of the wormhole starts increasing. Fig. \ref{fig:M2cpl} shows that for accretion of HDE the mass of wormhole shows decaying pattern during the quintessence phase and after the universe enters into phantom phase the mass of the wormhole starts increasing. Fig. \ref{fig:M3cpl} shows that the accretion of HDE onto wormhole leads to decay of the wormhole mass up to a finite value and in the phantom phase the mass increases indefinitely.

\subsection{JBP parameterization}
\begin{figure}[h]
\begin{minipage}{16pc}
\includegraphics[width=16pc]{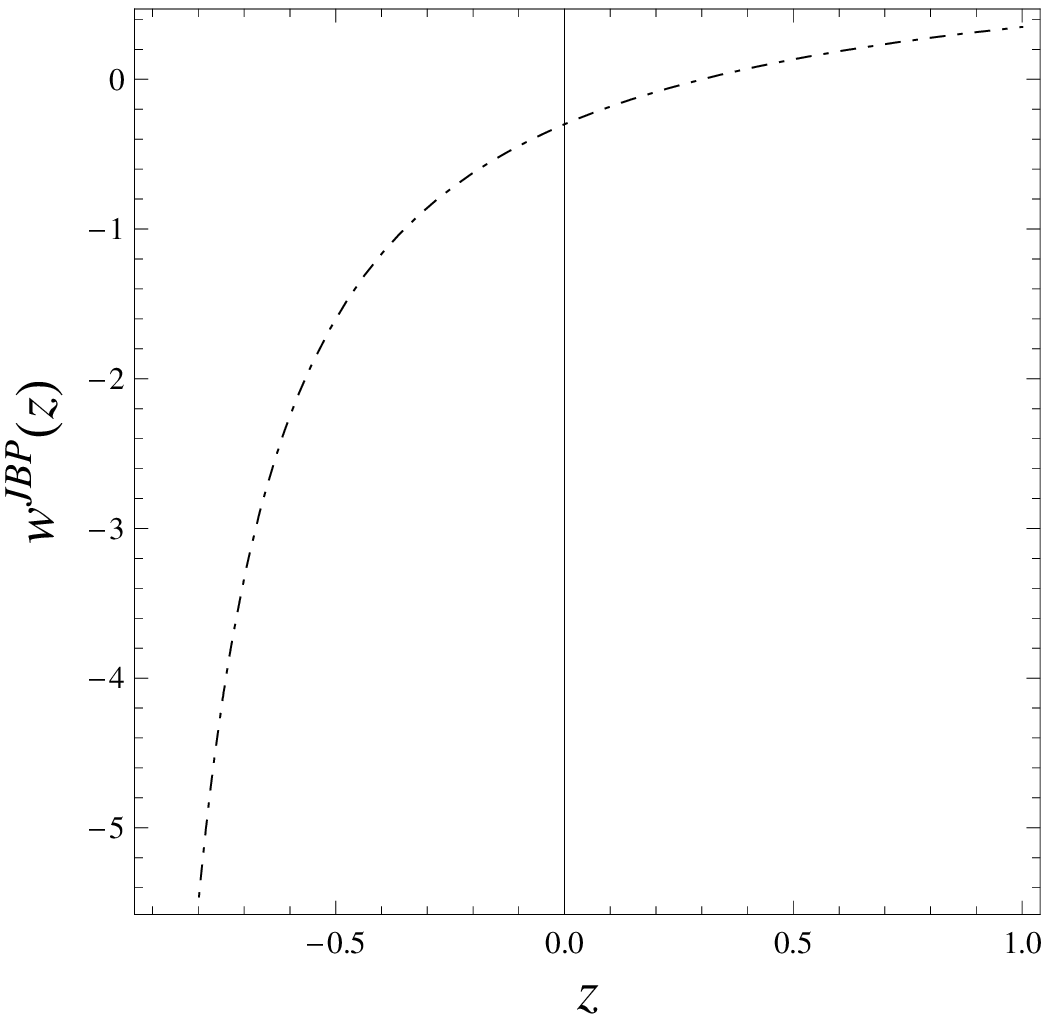}
\caption{\label{fig:eosjbp} EoS for JBP parameterization with $w_0=-0.3,~w_1=1.3$.}
\end{minipage}\hspace{3pc}%
\begin{minipage}{16pc}
\includegraphics[width=16pc]{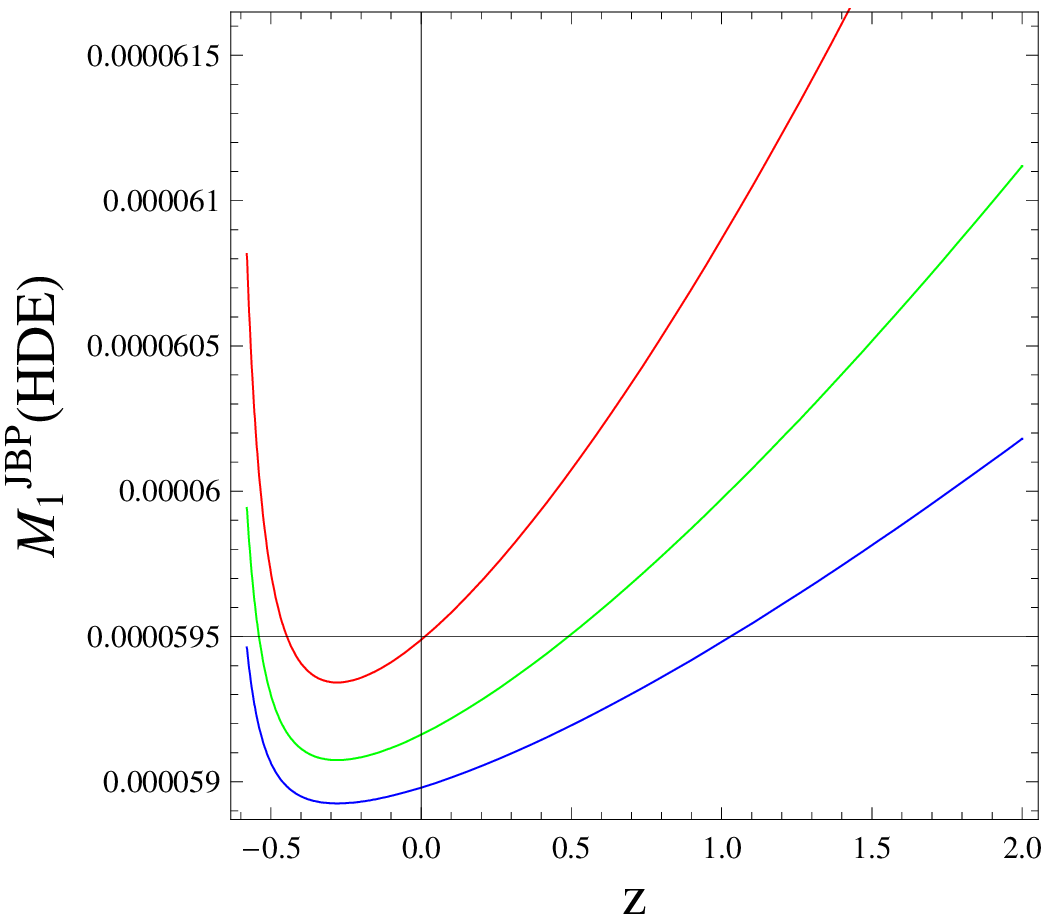}
\caption{\label{fig:M1jbp} Variations of wormhole mass $M_1$ (Eq. (\ref{M1})) against redshift $z=a^{-1}-1$ for HDE using JBP parameterization.}
\end{minipage}\hspace{3pc}%
\begin{minipage}{16pc}
\includegraphics[width=16pc]{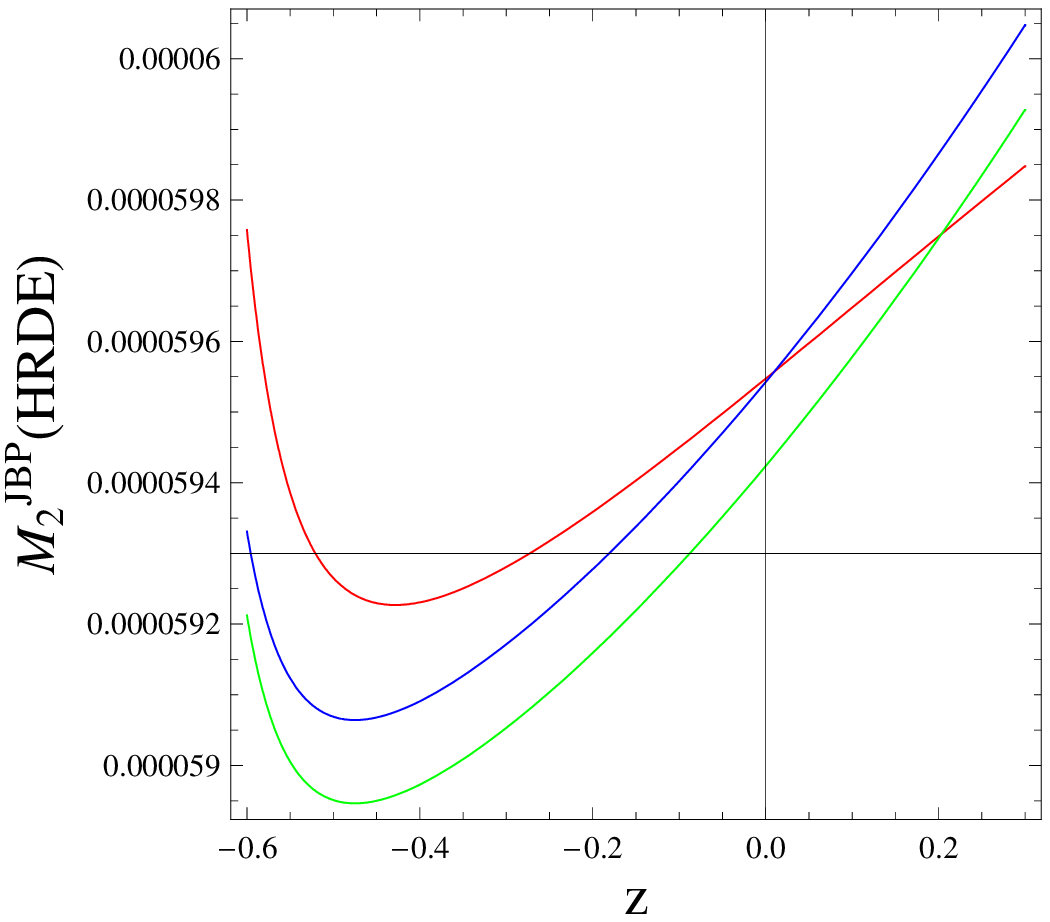}
\caption{\label{fig:M2jbp}Variations of wormhole mass $M_2$ (Eq. (\ref{M2})) against redshift $z=a^{-1}-1$ for HRDE using JBP parameterization.}
\end{minipage}\hspace{3pc}%
\begin{minipage}{16pc}
\includegraphics[width=16pc]{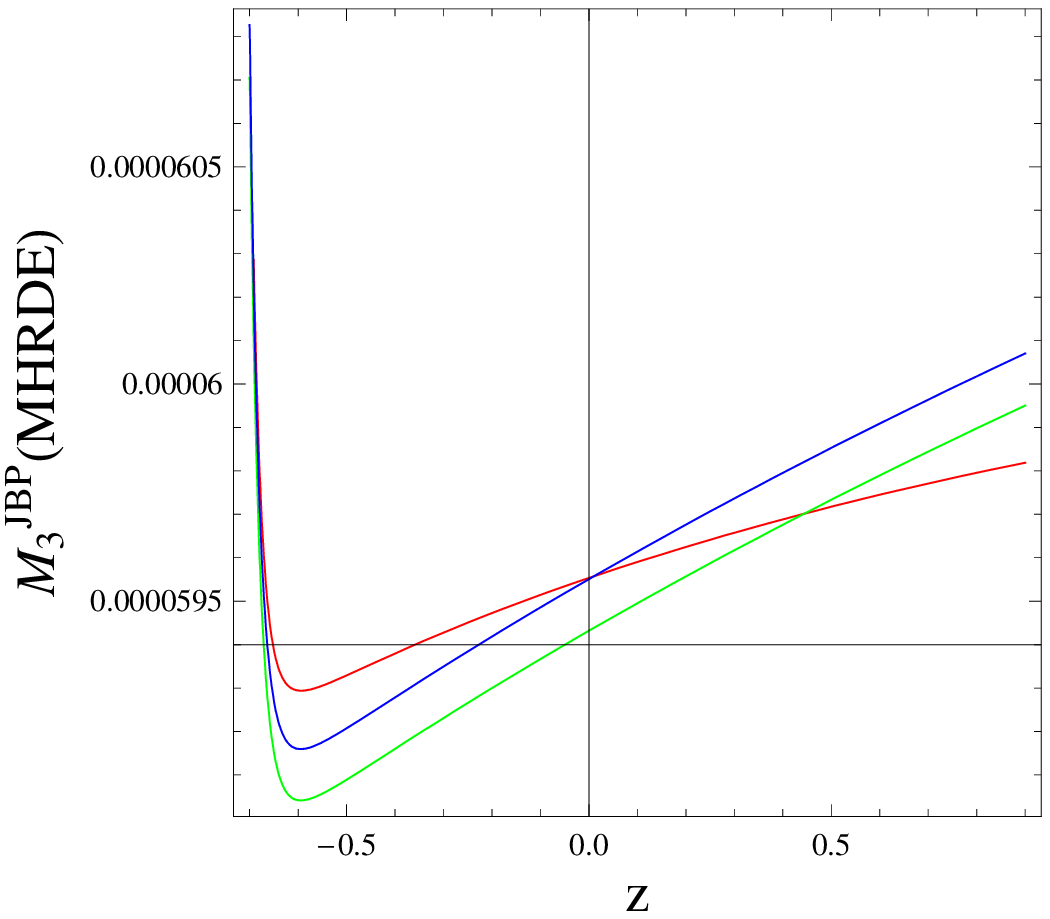}
\caption{\label{fig:M3jbp}Variations of wormhole mass $M_3$ (Eq. (\ref{M3})) against redshift $z=a^{-1}-1$ for MHRDE using JBP parameterization.}
\end{minipage}\hspace{3pc}%
\end{figure}
Jassal et al \cite{JBP} used a parameterization of the form
\begin{equation}\label{JBP}
w^{JBP}(z)=w_0+w_1 \frac{z}{(1+z)^2}
\end{equation}
that is obtainable by putting $m=2$ in Family II. Since $a=(1+z)^{-1}$, the conservation equation takes the form
\begin{equation}
(1+z)\frac{d\rho}{dz}=3\rho\left(w_0+w_1 \frac{z}{(1+z)^2}\right)
\end{equation}
Solving the above differential equation we obtain
\begin{equation}\label{rhojbp}
\rho(z)=\rho_0(1+z)^{3(1+w_0)}\exp\left[\frac{3 w_1 z^2}{2(1+z)^2}\right]
\end{equation}
Combining Eq. (\ref{hde}), and (\ref{rhojbp}) i.e. considering $\rho_{HDE}=\rho$ in Eq. (\ref{rhojbp}) we get differential equation that is not having any exact solution. Thus, we solve them numerically and use the solution in Eq. (\ref{mass}) to have a plot of $M$ in Fig. \ref{fig:M1jbp}. Similarly, numerical technique is applied for HRDE and MHRDE also and the wormhole masses are presented in Figs. \ref{fig:M2jbp} and \ref{fig:M3jbp} . 

In Fig. \ref{fig:eosjbp} we have plotted the EoS parameter for $w_0=-0.3,~w_1=1.3$ and observed a transition from quintessence to phantom phase at $z\approx-0.4$. As the accretion of HDE occurs on the wormhole (Fig. \ref{fig:M1jbp}), its mass is decreasing as long as the universe is in quintessence phase and as soon as it enters the phantom phase the mass of the wormhole starts increasing. Fig. \ref{fig:M2jbp} shows that for accretion of HRDE the mass of wormhole shows decaying pattern during the quintessence phase and after the universe enters into phantom phase the mass of the wormhole starts increasing. Fig. \ref{fig:M3jbp} shows similar pattern to that of HRDE and here also the accretion of MHRDE onto wormhole leads to decay of the wormhole mass up to a finite value and in the phantom phase the mass increases indefinitely.
\subsection{Efstathiou parameterization}
\begin{figure}[h]
\begin{minipage}{16pc}
\includegraphics[width=16pc]{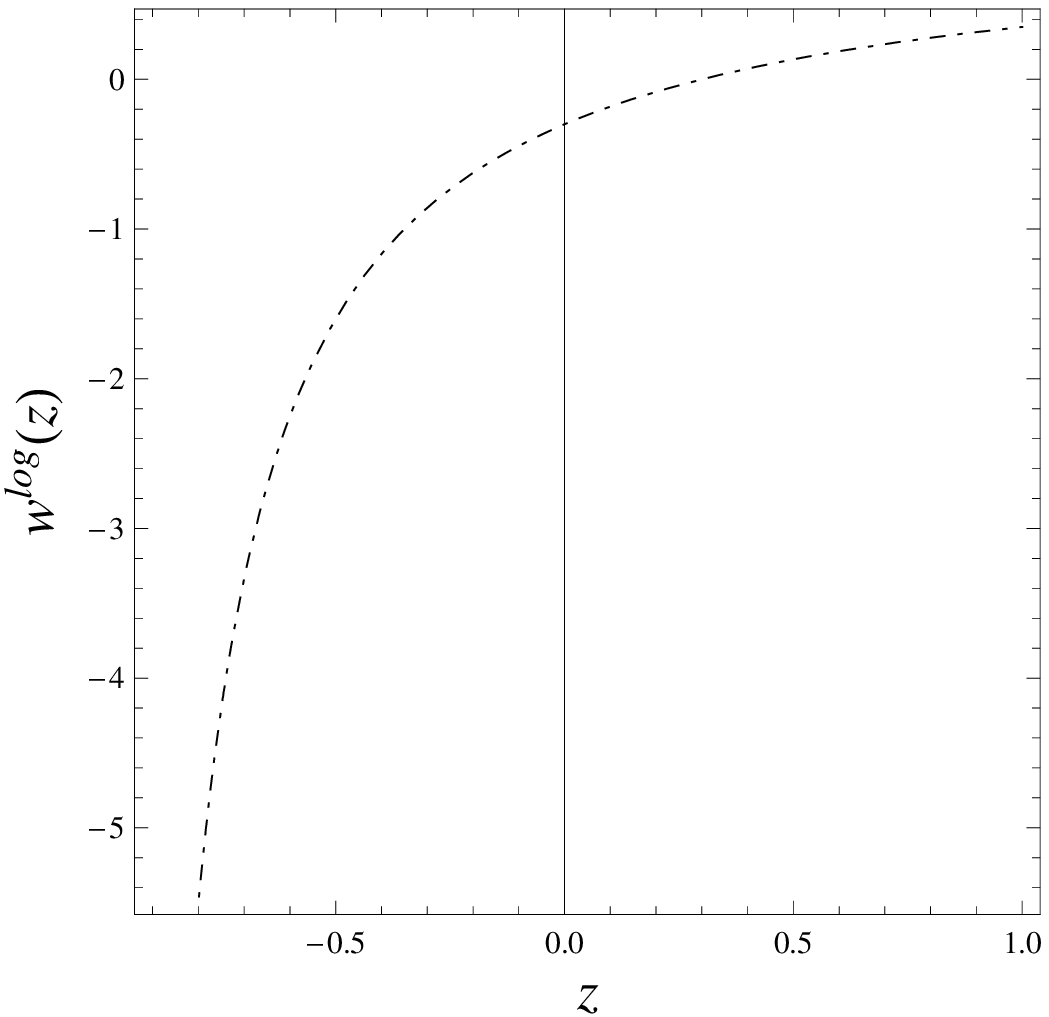}
\caption{\label{fig:eoslog} EoS for Efstathiou parameterization with $w_0=-0.3,~w_1=1.3$.}
\end{minipage}\hspace{3pc}%
\begin{minipage}{16pc}
\includegraphics[width=16pc]{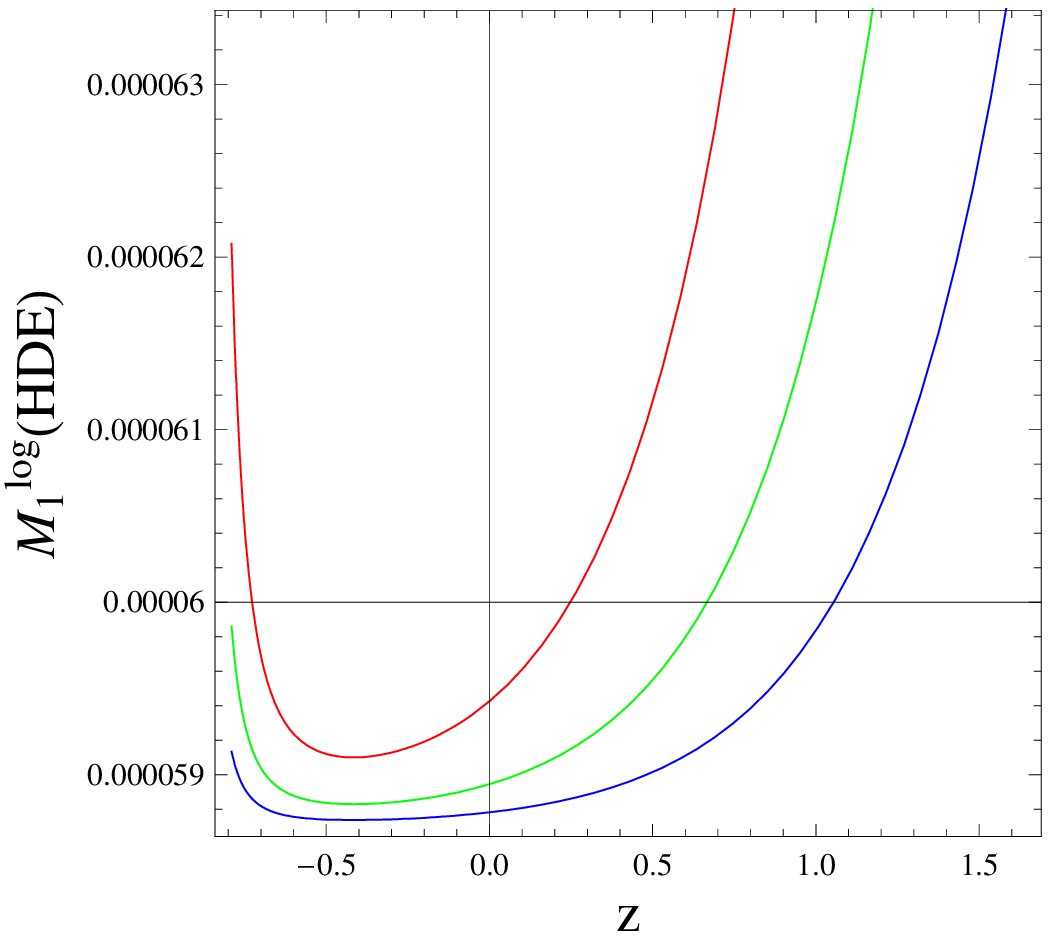}
\caption{\label{fig:M1log} Variations of wormhole mass $M_1$ (Eq. (\ref{M1})) against redshift $z=a^{-1}-1$ for HDE using Efstathiou parameterization.}
\end{minipage}\hspace{3pc}%
\begin{minipage}{16pc}
\includegraphics[width=16pc]{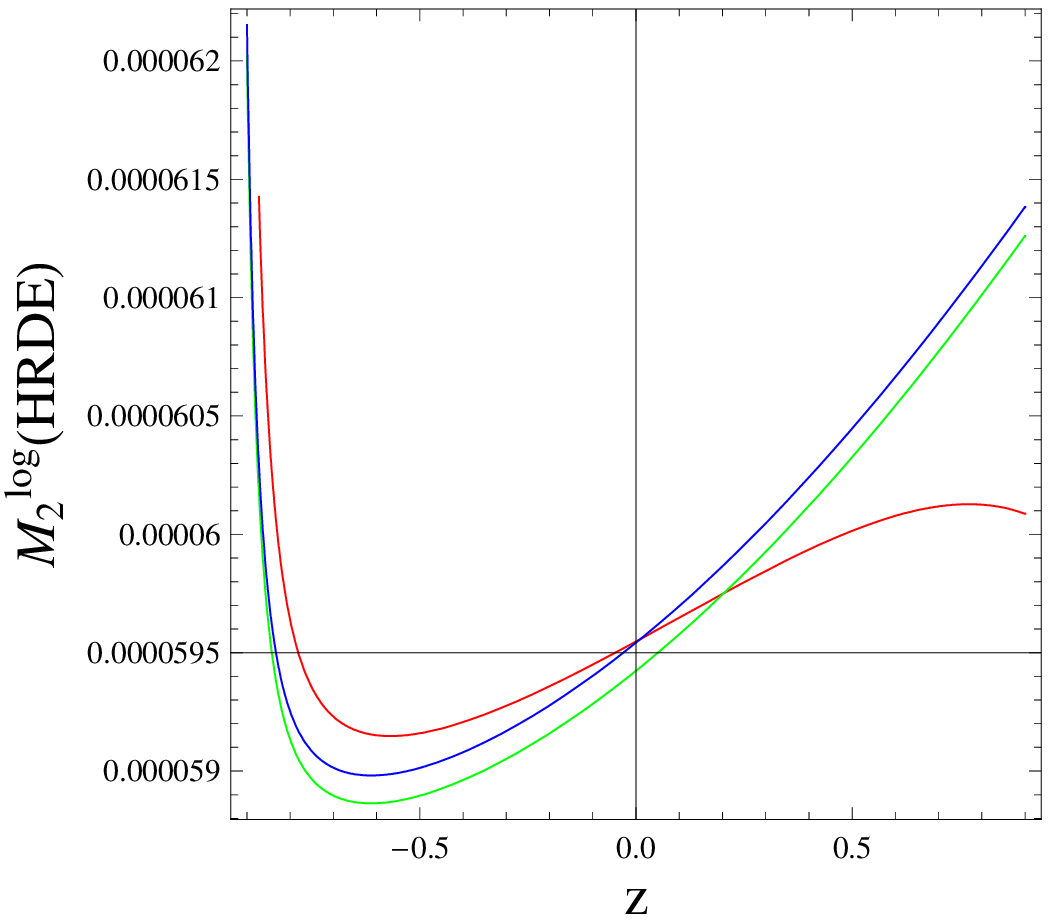}
\caption{\label{fig:M2log}Variations of wormhole mass $M_2$ (Eq. (\ref{M2})) against redshift $z=a^{-1}-1$ for HRDE using Efstathiou parameterization.}
\end{minipage}\hspace{3pc}%
\begin{minipage}{16pc}
\includegraphics[width=16pc]{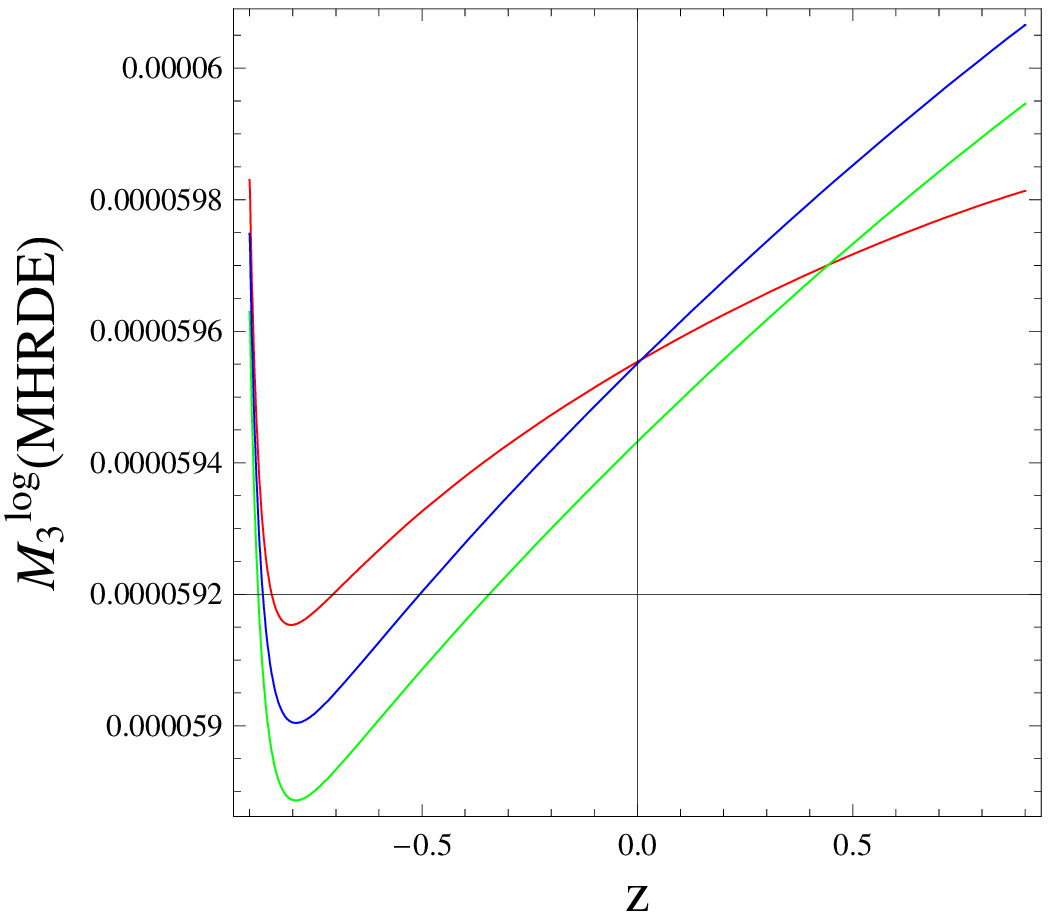}
\caption{\label{fig:M3log}Variations of wormhole mass $M_3$ (Eq. (\ref{M3})) against redshift $z=a^{-1}-1$ for MHRDE using Efstathiou parameterization.}
\end{minipage}\hspace{3pc}%
\end{figure}

Efstathiou  proposed the following parameterization for EoS \cite{Efstathiou}:
\begin{equation}\label{logeos}
w^{log}(z)=w_0+w_1\ln(1+z)
\end{equation}
For this parameterization we have
\begin{equation}
\rho=\rho_0~e^{\frac{3}{2}w_1 (\ln[1+z])^2} (1+z)^{3+3 w_0}
\end{equation}
For $\rho=\rho_{HDE}$, $\rho=\rho_{HRDE}$ and $\rho_{MHRDE}$ we get differential equations that are solved numerically and wormhole masses are plotted in Figs. \ref{fig:M1log}, \ref{fig:M2log} and \ref{fig:M3log} respectively.

In Fig. \ref{fig:eoslog} we have plotted the EoS parameter for $w_0=-0.3,~w_1=1.3$ and observed a transition from quintessence to phantom phase at $z\approx-0.4$. As the accretion of HDE occurs on the wormhole (Fig. \ref{fig:M1log}), its mass is decreasing as long as the universe is in quintessence phase and as soon as it enters the phantom phase the mass of the wormhole starts increasing. Fig. \ref{fig:M2log} shows that for accretion of HRDE the mass of wormhole shows decaying pattern during the quintessence phase and after the universe enters into phantom phase the mass of the wormhole starts increasing. Fig. \ref{fig:M3log} shows similar pattern to that of HRDE and here also the accretion of MHRDE onto wormhole leads to decay of the wormhole mass up to a finite value and in the phantom phase the mass increases indefinitely.

\section{Concluding remarks}
The present paper reports a study of accretion of three holographic dark energies, namely, holographic dark energy (HDE) with density $\rho_{HDE}=3c^2M_p^2H^2$ ; holographic Ricci dark energy with density $\rho_{HRDE}=3c^2(\dot{H}+2H^2)$, and modified holographic Ricci dark energy (MHRDE) with density  $\rho_{MHRDE}=\frac{2}{\alpha-\beta}\left(\dot{H}+\frac{3\alpha}{2}H^2\right)$ ($\alpha$ and $\beta$ are free constants), onto Morris-Thorne wormhole. This work has been carried out in two phases. In first phase of the study we assumed a scale factor of the from $a=a_0t^m, ~(m>0)$ leading to $H=m/t$ leading to $\dot{H}<0$. In the remaining part work we have followed the procedure of studying dark energy accretion onto wormhole already adopted by \cite{ujjal1,pedro}. It has been emphasized in \cite{ujjal1} that mass of the wormhole is a dynamical quantity, so the nature of the mass function is important in our wormhole model for different dark energy filled universe. Mass of the wormhole depends upon $p+\rho$ and hence on $\dot{H}$. Negative sign of $\dot{H}$ is contributed by $w>-1$ i.e. quintessence. We have obtained the wormhole mass for accretion of the above three dark energies in a universe characterized by power-law form of scale factor. We have observed that for the said accretion processes the mass of the wormhole is decaying with the evolution of the universe. Ref. \cite{Babichev} has shown that although the mass of black hole decreases due to phantom energy accretion, the mass of wormhole increases due to phantom
energy accretion, which is the opposite behaviour of black hole mass. Also, \cite{podaro} had shown that in the dual phantom generalised Chaplygin gas wormholes decrease their size when the scale factor increases. Here, considering three holographic dark energy models we observe that for increase in the scale factor the mass of wormhole is decaying. Hence, this observation is consistent with \cite{podaro}. 

In the next phase of the study we have adopted three dark energy parameterization schemes $w^{CPL}(z)=w_0+w_1 \frac{z}{1+z},~~w^{JBP}(z)=w_0+w_1 \frac{z}{(1+z)^2},~~w^{log}(z)=w_0+w_1\ln(1+z)$ proposed in \cite{CPL2,JBP,Efstathiou}. In this phase we have not assumed any form of the scale factor and solved the conservation equations to get the dark energy densities under the three parameterization schemes. Subsequently, considering this dark energy as HDE, HRDE and MHRDE we have reconstructed the Hubble parameter in each case and this way we obtained wormhole masses when the dark energies are accreting on the wormhole. We have taken $w_0=-0.3,~w_1=1.3$. In all parameterization cases we observed that during the quintessence phase of the universe the mass of wormhole is decreasing with $z$ and the mass is increasing in the phantom phase of the universe. Ref. \cite{suskov} explicitly demonstrated that the phantom energy can support the existence of static wormholes. Hence, our observation that the holographic dark energies, when accreting on wormhole, leading to an increase in the wormhole mass in the phantom phase of the universe is in agreement with \cite{suskov}. Also, the study is in agreement with \cite{ujjal1}, who showed that accretion of two types of Chaplygin gas onto Morris-Thorne Wormhole lead to increase in wormhole mass in the phantom phase of the universe.

\subsection{Acknowledgement}
 Surajit Chattopadhyay received financial support from Department of Science and Technology, Govt. of India, under Project Grant No. SR/FTP/PS-167/2011 is duly acknowledged.

\end{document}